\documentclass[10pt]{article}
\usepackage[letterpaper]{geometry}
\usepackage{hicss}
\usepackage{times}
\usepackage[none]{hyphenat}
\usepackage{url}
\usepackage{latexsym}
\usepackage{graphicx}
\usepackage{subcaption}
\usepackage{caption}
\graphicspath{{images/}}
\usepackage[
    style=ieee,
    maxnames=1,    % Show only one author before 'et al.' in citations
    minnames=1,    % Ensure 'et al.' is used when there are multiple authors
    maxbibnames=2 % (Optional) Show up to 99 authors in the bibliography
]{biblatex}
\usepackage{enumitem}

\usepackage{indentfirst}
\usepackage{xcolor}
\usepackage{amsmath}
\usepackage{graphicx}
\usepackage{float}
\usepackage{algorithmic}
\usepackage[linesnumbered,ruled]{algorithm2e}
\usepackage{array}
\usepackage{tabularx}
\usepackage{optidef} % For optimization problem formatting
\newcommand{\Hquad}{\hspace{0.5em}} 
\usepackage{booktabs}
\usepackage{amssymb}
\usepackage{multirow}
\usepackage{makecell}
\usepackage{threeparttable}
\usepackage{indentfirst}
\usepackage{hyperref}
\usepackage{enumitem,kantlipsum}
\newcommand{\iv}{$IV$}
\usepackage{xcolor}

\usepackage{hyperref}
\usepackage{multicol}
\usepackage{fancyhdr}
\title{\textbf{Distributed Primal-Dual Interior Point Framework for Analyzing Infeasible Combined Transmission and Distribution Grid Networks}}
\author{
    \begin{minipage}{0.45\textwidth}
        \centering
        Muhammad Hamza Ali \\
        University of Vermont \\
        \underline{Muhammad-Hamza.Ali@uvm.edu}
    \end{minipage}%
    \hspace{0.1\textwidth} % Adjusts space between authors
    \begin{minipage}{0.45\textwidth}
        \centering
        Amritanshu Pandey \\
        University of Vermont \\
        \underline{amritanshu.pandey@uvm.edu}
    \end{minipage}
    \vspace{-12mm}
}
\date{}
\vspace{-8mm}
\begin{document}
\maketitle
\thispagestyle{fancy} % for pre-print version
\begin{abstract}
\vspace{-1.0em}
\noindent The proliferation of distributed energy resources has heightened the interactions between transmission and distribution (T\&D) systems, necessitating novel analyses for the reliable operation and planning of interconnected T\&D networks. 
A critical gap is an analysis approach that identifies and localizes the weak spots in the combined T\&D networks, providing valuable information to system planners and operators. 
The research goal is to efficiently model and simulate infeasible (i.e. unsolvable in general settings) combined positive sequence transmission and three-phase distribution networks with a unified solution algorithm. We model the combined T\&D network with the equivalent circuit formulation. 
To solve the overall T\&D network, we build a Gauss-Jacobi-Newton (GJN) based distributed primal dual interior point optimization algorithm capable of isolating weak nodes. We validate the approach on large combined T\&D networks with 70k+ T and 15k+ D nodes and demonstrate performance improvement over the alternating direction method of multipliers (ADMM) method.
\end{abstract}
%\subsubsection*{Keywords:} 
\vspace{2pt}
\noindent \textbf{Keywords:} Combined T\&D networks, distributed optimization, grid planning, infeasibility analysis, primal-dual interior point.
\vspace{-0.7em}
\section{Introduction}
\vspace{-1.3em}
\noindent Integration of distributed energy resources (DERs) like photovoltaics, battery storage systems, heat pumps, and electric vehicles is transforming the electric grid architecture and making it more heterogenous, distributed, and sustainable \cite{aguero2018grid}. 
Although the changing grid architecture offers many benefits, recognizing its inherent complexities is critical, especially in the context of co-dependency between the transmission and distribution (T\&D) grid operation and planning.

The fundamental basis for power systems planning and operation lies in the power flow (PF) studies, which establish a relationship between complex voltage phasors and nodal power injections, encompassing both active and reactive powers at various busses within the network. 
Traditionally, for bulk power system (BPS) operations and planning, grid planners aggregate the distribution feeders as a single entity (e.g., PQ or ZIP model) at the transmission-distribution (T\&D) interface. 
In this paradigm, the distribution system, being inherently passive, assumes the role of an energy requester from the transmission network. 
As a result, the transmission BPS resources manage voltage support, energy balance, and various essential adjustments for the distribution network. %\textcite{mohseni2020transmission,nardelli2014models}.  
A recent directive from the Federal Energy Regulatory Commission (FERC), Order No. 841 \cite{FERC841}, mandates the establishment of a participation model for energy storage services with a capacity exceeding 0.1 MW in the wholesale market, regardless of their location. 
This regulatory shift aligns with the growing significance of DERs. 
It emphasizes the need to explore the combined operation of T\&D networks, especially in areas like Vermont, U.S., where DERs can constitute most of the state's net generation during sunny days. 
During these scenarios, the grid observes backfeed at many T\&D interfaces in the region.
To operate and plan for such grid scenarios, we need to analyze and optimize combined T\&D networks.

% \vspace{-0.5em}
There are many classes of combined T\&D problems, including time-domain and steady-state analyses \cite{sun2014master, huang2016integrated, aristidou2014dynamic}. 
We focus on steady-state combined T\&D models, specifically identifying and simulating the infeasible networks (i.e., these networks have no feasible AC solution without adjustments).
In the context of high renewable penetration, this approach enables system planners to identify weak locations within T\&D networks precisely. 
%Moreover, it allows us to study scenarios where resources in the transmission drive the reliable operation of distribution network and vice-versa.
The need is even more apparent as the operational dynamics of distribution networks evolve beyond passive roles. 
Today's distribution networks experience more active control mechanisms, such as inverter control, photovoltaic (PV) curtailment, and volt-var control etc. 
These active elements introduce complexities that influence the reliability of tightly coupled T\&D systems \cite{miller2018impacts}.
%The need is even more apparent when the reliability of transmission and distribution networks during control actions is interdependent, necessitating an understanding of the operational state of both distribution and transmission systems.
%In addition to this, these studies play a pivotal role in bridging the gap between Transmission System Operators (TSO) and Distribution System Operators (DSO), facilitating effective coordination that underpins both preventive and corrective measures for various challenges, such as voltage collapse or identifying regions with real or reactive power deficiencies. 
An illustrative example of this significance is from 2013 when the transmission system operator in PJM collaborated with the Sturgis, Michigan distribution system to prevent a blackout, utilizing 6MW from the distribution grid \cite{interconnection2014analysis}.
In a similar event in 2022, roughly 2400 homeowners pushed about 16.5 MW of power back to the grid to test Tesla's virtual power plant capabilities \cite{teslavpp}.
%%%%%%%%%%%%%%%%%%%%%%%%%%%%%%%%%%%%%%%%%%%%%%%%%%%%%%%%%%%%%%%%%%%%%%%%%%%%%%%%%%%%%%%%%

The state-of-the-art combined T\&D algorithms typically fall into two categories: co-simulation \cite{kalsi2013integrated, anderson2014gridspice} and co-modelling approaches \cite{Pandey_CombinedTD, sun2014master}. 
The co-simulation approach employs separate tools to model transmission and distribution components with distinct formulations. 
Recent developments include tools like FNCS\footnote{Framework for network co-simulation.} 
\cite{ciraci2014fncs}, IGMS\footnote{Integrated grid modelling system} \cite{palmintier2016igms}, HELICS\footnote{Hierarchical engine for large scale infrastructure \& co-simulation.} \cite{8064542}, designed to facilitate T\&D co-simulation. 
A co-simulation methodology is described in \cite{kalsi2013integrated}, where PowerWorld/PSSE solved the transmission system, and GridLAB-D/OpenDSS solved the distribution grid.  In most of these methods, information exchange occurs over the communication layer. 
However, a significant drawback is that users must devise appropriate interfacing techniques for diverse simulation tools, and the overall approach fails if any sub-problem cannot yield a solution. 
Finally, in the context of challenging T\&D optimization problems, interfacing between T and D solvers involves only primal variables, and convergence guarantees are difficult to obtain. 

In the co-modeling approach, as highlighted in \cite{Pandey_CombinedTD}, merely addressing the power flow is insufficient for offering meaningful planning insights as simulation can fail due to many reasons, \textit{e.g.}, misrepresentation of a real physical system, lack of AC feasibility under given operating conditions, poor choice of initial conditions, etc.
System planners must understand the root cause of power flow simulation failure.
\textit{E.g.} consider a scenario where a segment of the transmission network becomes infeasible due to excessive PV exports from the interconnected distribution network. 
In this instance, curtailing the PV within the distribution network can restore feasibility to the overall system. Therefore, we need techniques to detect infeasible T\&D networks and offer valuable information on the cause and remedial action for failure.
\begin{table}[htb!]
    \centering
    \caption{Classification of Infeasibilities}
    \begin{tabularx}{\columnwidth}{|X|X|X|X|}
    \hline
    \multicolumn{1}{|c|}{Network} & \multicolumn{3}{c|}{Infeasibility type}
    \\ \hline \hline
    $\mathcal{I}^{T}$ & $I^{inf}$ & $S^{inf}$  & $Z^{inf}$  \\
    $\mathcal{I}^{D}_{\Omega}$ & $I^{inf}_{\Omega}$ & $S^{inf}_{\Omega}$  & $Z^{inf}_{\Omega}$ \\ \hline
    \end{tabularx}
    \label{tab:1}
\end{table}
\vspace{-1em} % Adjust this value as needed

To address current industry needs and technology gaps, we propose a combined T\&D infeasibility analysis framework that identifies and locates weak spots\footnote{areas or nodes in either transmission or distribution networks that require active/reactive power support.} and offers tailored remedies for individual infeasible scenarios. 
We consider key challenges: stakeholders from different T\&D utilities often resist sharing their internal network data, and solving iteration matrices with millions of nodes is computationally not tractable under a single machine, single memory computing framework \cite{pandey2024combined}. 
Therefore, we develop a \textit{Gauss-Jacobi-Newton (GJN) based distributed primal dual interior point (D-PDIP) optimization method}, which addresses both privacy and computing challenges.
We show that the D-PDIP algorithm can outperform the alternating direction method of multipliers-based distributed algorithms as it considers second-order information for dual variable updates. The layout of the combined T\&D network is shown in Fig (\ref{fig:1}), and we can have an infeasibility source at every node on T and D networks or a subset of them. 
The available types of infeasibility sources for the different remedial actions are highlighted in Table \ref{tab:1}. 
The choice of network equations in our combined model is based on Kirchoff’s Current Law (KCL), where the infeasibility sources offset any mismatch in the equations. 
The distributed optimization approach in Section (\ref{opti}) minimizes the norm of the infeasibility sources to optimally compensate for the mismatches while considering the physical limit on node voltages and line flow.
Our approach has the following features:  
\vspace{-2mm}
\begin{enumerate}[leftmargin=*, noitemsep]
    \item \textbf{Robustness:} Model and simulate \textit{infeasible} combined T\&D networks within the same solution algorithm independent of the choice of initial or network conditions.
    \item \textbf{Scalability and Data-privacy:} Develop a Gauss-Jacobi-Newton based \textit{distributed} primal-dual interior point algorithm that minimizes data communication between entities and solves large networks ($>$80k nodes).
    \item \textbf{Generality:} Apply the proposed algorithm to various practical case-studies requiring analyses of tightly coupled combined T\&D networks.
\end{enumerate}
\vspace{-4mm}

\section{Preliminaries: \iv{}-based Transmission and Distribution Modeling}
\begin{figure*}[htbp]
    \centering
\includegraphics[width=0.93\textwidth,height=10\textwidth,keepaspectratio]{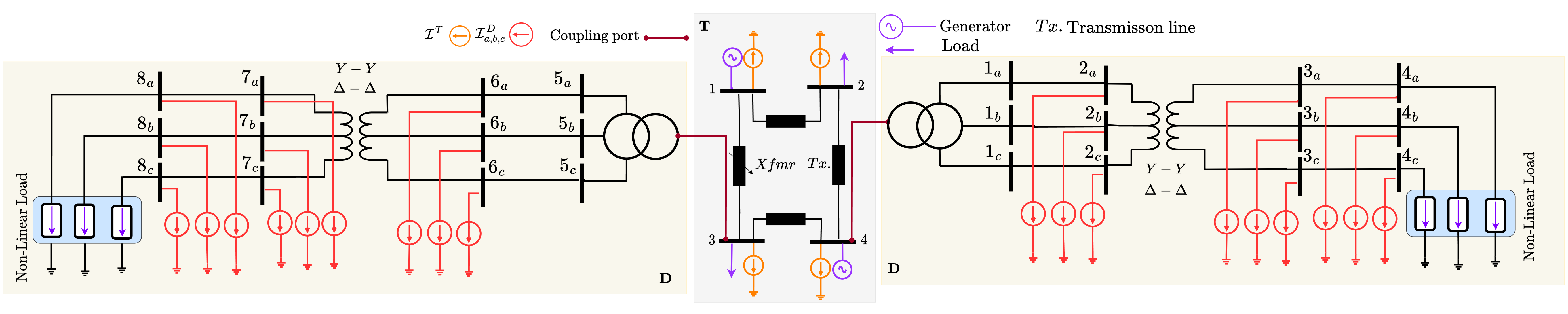}
    \caption{Illustration of combined positive sequence transmission network and three-phase distribution networks with infeasibility sources at each node.}
    \label{fig:1}
\end{figure*}

\vspace{-1.0em}
\noindent We employ the \iv{}-based equivalent circuit formulation (ECF) approach to model and study the infeasible combined T\&D networks. In ECF, \cite{8424869}, all network elements (line, shunts, transformer, etc.) except generator/load injections represent linear \iv{} relationships. Positive sequence \iv{} relationships model transmission networks, assuming balanced operation, whereas explicit three-phase \iv{} relationships model unbalanced distribution networks. A coupling port, as shown in Fig. (\ref{fig:1}), facilitates an interface between T and D sub-models and allows information exchange between the two networks at the point of interconnection (POI). 
\subsection{Positive-sequence Transmission Network} 
\vspace{-0.5em}
\noindent We represent the positive sequence transmission network with an undirected graph $\mathcal{G}^{T}(\mathcal{N}^{T}, \mathcal{E}^{T})$. The \iv{} relationships for node $\mathcal{N}^{T}$ and edge $\mathcal{E}^{T}$ elements model the network physics with KCL, which requires that positive sequence currents at each node $n^T \in \mathcal{N}^{T}$ sum to zero. Equations \eqref{eq_1} through \eqref{eq_7} formulate this relationship with major simplifications for readability, where $\Phi=1$ indicates positive sequence parameters and states. Node complex voltage states are represented in rectangular coordinates as real $(V^R)$ and imaginary $(V^I)$ voltages. $G_{ij}$ and $B_{ij}$ correspond to the admittance and susceptance of edge elements between nodes $i$ and $j$. The terms $I_{i}^{R}$ and $I_{i}^{I}$ represent the net current injection at node $i$ from loads and generators.

\subsection{Three-phase Distribution Network}
\vspace{-0.5em}
\noindent Similar to transmission networks, the AC network physics of the distribution grid model (denoted as $\mathcal{G}^D(\mathcal{N}^D, \mathcal{E}^D)$) is presented with major simplifications in equations \eqref{eq_1} through \eqref{eq_7}. 
Here, $\Phi$ represents the set of connected phases on each node ($n_D \in \mathcal{N}^{D}$), with a maximum of three phases ($\Phi = \{a, b, c\}$). The symbol $\Gamma$ encapsulates the self and mutual coupling between all the combinations of phases. The models include various three-phase transformers, shunts, switches, and fuses, but these are omitted for simplicity and readability. %Mathmatically, the formulation is defined in \eqref{eq_1} to \eqref{eq_6}, $i \in \mathcal{N}$ and $\Omega \in \Phi$. 
%\vspace{-0.5em}
\begin{subequations}
\vspace{-0.8em}
\setlength{\jot}{0.2em} 
\begin{alignat}{1}
    & \forall i \in \mathcal{N}, \forall \Omega \in \Phi: \nonumber  \\
    \small 
    & \sum_{j \in \mathcal{N}} \sum_{\Gamma \in \Phi} \left( G_{ij,\Omega\Gamma} V_{ij,\Gamma}^{R} - B_{ij,\Omega\Gamma} V_{ij,\Gamma}^{I} \right) + I_{i,\Omega}^{R} =0  \label{eq_1}\\ 
    &\small \sum_{j \in \mathcal{N}} \sum_{\Gamma \in \Phi} \left( G_{ij,\Omega\Gamma} V_{ij, \Gamma}^{I} + B_{ij,\Omega\Gamma} V_{ij,\Gamma}^{R}\right) + I_{i,\Omega} ^{I} = 0
    \label{eq_2} \\ 
    &I_{i,\Omega} ^{R} - \cfrac{P_{i,\Omega} V_{i,\Omega}^{R} + Q_{i,\Omega} V_{i,\Omega}^{I}}{(V_{i,\Omega}^{R})^2 + (V_{i,\Omega}^{I})^2} = 0 \label{eq_3} \\
    &I_{i,\Omega} ^{I} - \cfrac{P_{i,\Omega} V_{i,\Omega}^{I} - Q_{i,\Omega} V_{i,\Omega}^{R}}{(V_{i,\Omega}^{R})^2 + (V_{i,\Omega}^{I})^2} = 0 \label{eq_4} \\
    &P_{i, \Omega} - \sum_{l \in \mathcal{L}_i} P^l_{i,\Omega} + \sum_{g \in G_i} P^g_{i,\Omega} = 0  \label{eq_5}
\end{alignat}
\begin{alignat}{1}
    &Q_{i,\Omega} - \sum_{l \in \mathcal{L}_i}Q^l_{i,\Omega} + \sum_{g \in G_i}Q^g_{i,\Omega} = 0 \label{eq_6} \\
    &(V_{i, \Omega}^{R})^{2} + (V_{i, \Omega}^{I})^{2} - \hat{V}_{i, \Omega}^2 = 0 \quad \forall i \in \mathcal{N} \backslash PQ \label{eq_7}
\end{alignat}
\end{subequations}
\vspace{-1.5em}
\subsection{Infeasibility Analysis}
\noindent Prior works have focused on identifying and solving the infeasible positive sequence transmission networks. \cite{overbye1994power} introduced a method to identify unsolvable power flow networks by introducing \textit{missing power} in power mismatch equations.   \cite{jereminov2020evaluating}, \cite{li2020lasso} introduced a similar concept by adding infeasibility current sources within the ECF formulation. 
%The methods of detecting the infeasibility in the positive sequence can not be applied to the distribution system due to non-controllable customers and unbalance between the phases. 
For a three-phase distribution network, \cite{foster2022three} discusses the infeasibility problem with \textit{L-1} regularization to introduce the sparsity in the solution vector. 
\cite{zamzam2016beyond} applies a similar concept as \cite{overbye1994power} but in three-phase distribution grids. 
\begin{subequations}
\begin{alignat}{2}
    & \mathbf{P}_{1}: \min_{X, I_{i}^{\inf}} \sum_{i \in \mathcal{N}} \sum_{\Omega \in \Phi} \lVert I^{\inf}_{i,\Omega} \rVert_{p}^{p} \label{inf_eq_1} \\ 
    &\text{subject to:} \nonumber \\
    & \mathcal{F}_{i,\Omega}(X) - I^{\inf}_{i, \Omega} = 0 \quad \forall i \in \mathcal{N},\ \forall \Omega \in \Phi \label{inf_eq_2} \\
    & \underline{X} \leq X \leq \overline{X} \label{inf_eq_3}
\end{alignat}
\end{subequations}
The \iv{}-based infeasibility formulation for T and D, independently, with ECF, is shown in $\mathbf{P_1}$ \eqref{inf_eq_1}-\eqref{inf_eq_3}. 
$\mathcal{F}(X)$ represents the power flow equations as shown in \eqref{eq_1}-\eqref{eq_7}, where current infeasibility source, $I^{\inf}_{\Omega}$, as described in \cite{jereminov2020evaluating}, \cite{li2020lasso}, and \cite{foster2022three} is added to each bus $i$ except the slack bus. 
$X$ is the vector of unknown variables, i.e $X \in \{X^{T}, X^{D}\}$. 
These methods address infeasibility for transmission or distribution separately and are developed for a centralized computing framework. 
In the combined T\&D context, incorporating additional current sources does not effectively isolate the exact cause of infeasibility, nor does it suffice to accurately devise preventive strategies in combined T\&D operations. 
Real-time combined T\&D networks can involve tens of millions of positive-sequence and three-phase nodes, surpassing the current capabilities of existing methods in the literature. 
They also have implicit privacy requirements, further necessitating the need for distributed approaches. 
Therefore, current approaches do not scale nor apply to combined T\&D analyses without loss of generality. We will target the gap in this paper.
\vspace{-0.2em}
\subsection{Coupling Port}
\begin{figure}[H]
    \centering
\includegraphics[width=0.9\columnwidth]{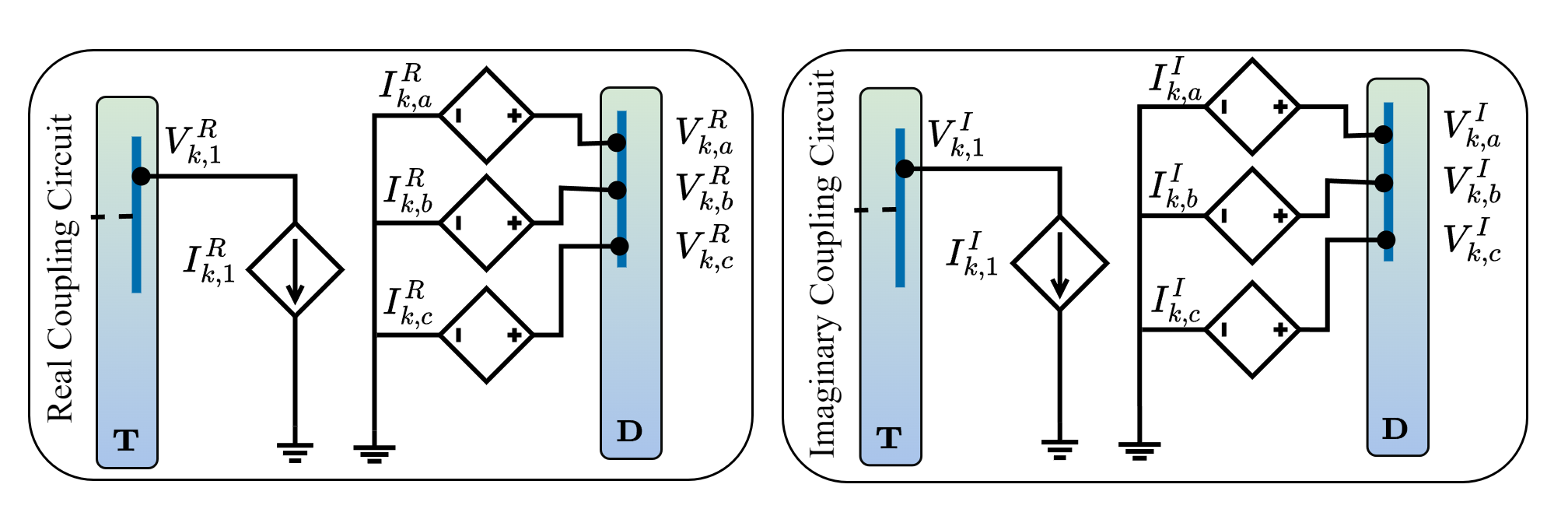}
     \caption{Real and imaginary circuit of coupling port.}
    \label{fig:5}
\end{figure}

\vspace{-0.5em}
\noindent We apply a circuit-theoretic approach to design a coupling port (with subscript $k$) that interconnects the positive sequence transmission network with the three-phase distribution network \cite{Pandey_CombinedTD}. 
The coupling port allows for the natural decomposition of T\&D subcircuits by replacing the controlled sources with externally updated independent sources.
The effect of the distribution network on the transmission side is represented by the positive-sequence current-controlled current sources $I^{R}_{k,1}$ and $I^I_{k,1}$. Conversely, voltage-controlled voltage sources ($V^{R}_{k,a}, V^{R}_{k,b}, V^{R}_{k,c}, V^{I}_{k,a}, V^{I}_{k,b}, V^{I}_{k,c}$) couple the distribution side voltages with the transmission POI voltages as shown in Fig. (\ref{fig:5}). 
\vspace{-0.5em}
The variable $\kappa$ is a normalizing constant given by the base current on the distribution side ($\kappa = S_{base}/V_{base}$), where $V_{base}$ is the nominal voltage of the distribution coupling node.
Assuming positive sequence components only on the transmission side, we use symmetrical components theory to derive the relationship \eqref{C_eq_1}.
\vspace{-6pt}
\begin{equation}
    \begin{bmatrix}
        {I}^{R}_{k,1} \\
        {I}^{I}_{k,1}
    \end{bmatrix} = \cfrac{1}{3\kappa}\begin{bmatrix}
        1 & 0 & \alpha & 0 & \alpha^{2} & 0 \\
        0 & 1 & 0 & \alpha & 0 & \alpha^{2} 
    \end{bmatrix}
    \begin{bmatrix}
        {I}^{R}_{k,a} \\
        {I}^{I}_{k,a} \\
        {I}^{R}_{k,b} \\
        {I}^{I}_{k,b} \\
        {I}^{R}_{k,c} \\
        {I}^{I}_{k,c}
     \end{bmatrix}
     \label{C_eq_1}
\end{equation}
Similarly, \eqref{couple_eq_2} represents the distribution sub-circuit voltages as a function of transmission voltages at POI.
\begin{equation}
\begin{bmatrix} 
            {V}^{R}_{k,a} \\
            {V}^{I}_{k,a} \\
            {V}^{R}_{k,b} \\
            {V}^{I}_{k,b} \\
            {V}^{R}_{k,c} \\
            {V}^{I}_{k,c}
         \end{bmatrix} = V_{base}\begin{bmatrix}
             1 & 0 \\
             0 & 1 \\
             \alpha^{2} & 0 \\
             0 & \alpha^{2} \\
             \alpha & 0 \\
             0 & \alpha 
         \end{bmatrix} \begin{bmatrix}
             {V}^{R}_{k,1} \\
             {V}^{I}_{k,1}
         \end{bmatrix}
         \label{couple_eq_2}
\end{equation}
In the (\ref{C_eq_1}) and (\ref{couple_eq_2}), the value of $\alpha$ is $\frac{2\pi}{3}$ radians.
These equations describe the relationship between a primal set of variables for the coupling port; in the optimization context, the relationship for the dual variables is derived in Section \ref{opti}.
\vspace{-0.7em}

%%%%%%%%%%%%%%% Need to add information about dual variables %%%%%%%%%%%%%%%%%%%
%%%%%%%%%%%%%%%%%%%%%%%%%%%%%%%%%%%%%%%%%%
\section{Combined T\&D Infeasibility Analysis}{\label{opti}}
\vspace{-0.5em}
\noindent To solve \textit{infeasible} combined T\&D networks, we first introduce a centralized approach $\mathbf{P_2}$. Next, we develop a decentralized method to enforce privacy needs and ensure scalability. In particular, we focus on enforcing performance and robustness in the distributed algorithm.
\vspace{-1.5em}
\begin{table}[t]
\small
\centering
\caption{Symbols and definitions \label{tab:notations}}
\begin{tabularx}{\columnwidth}{@{}lX@{}}
\toprule
\textbf{Symbol} & \textbf{Interpretation} \\ 
\midrule
$X^{T}$, $X^{D}_\Omega$ & Transmission and distribution state variables, $\forall \Hquad \Omega \in \{a,b,c\}$ \\
\midrule
$k \in K$ & Point of interconnections in the set of all interconnections between T and D \\
\midrule
$p$ & Choice of norm \\
\midrule
$X^{T}_{s}$, $X^{D}_{s,\Omega}$ & Transmission and distribution state variables in sub-problem $s \in S^{T},S^{D}$ and $\Omega \in \{a,b,c\}$ \\
\midrule
$\mathcal{I}_{s}^{T}$, $\mathcal{I}_{s,\Omega}^{D}$ & Vector of infeasibility source at T\&D in sub-problem $s \in S^{T}, S^{D}$, $\forall \Omega = \{a,b,c\}$ \\
\midrule
$X^{T,int}_{s}$, $X^{D,int}_{s,\Omega}$ & Internal variables in T\&D $\forall s \in S^{T}, S^{D}, \Hquad \Omega \in \{a,b,c\}$ \\
$X^{T,ext}_{s}$, $X^{D,ext}_{s,\Omega}$ & External variables in T\&D $\forall s \in S^{T}, S^{D}, \Hquad \Omega \in \{a,b,c\}$ \\
$\lambda^{T,int}_{s}$, $\lambda^{D,int}_{s,\Omega}$ & Dual internal variables in T\&D $\forall s \in S^{T}, S^{D}, \Hquad \Omega \in \{a,b,c\}$ \\
$\lambda^{T,ext}_{s}$, $\lambda^{D,ext}_{s,\Omega}$ & Dual external variables in T\&D $\forall s \in S^{T}, S^{D}, \Hquad \Omega \in \{a,b,c\}$ \\
\bottomrule
\end{tabularx}
\end{table}
\subsection{Centralized approach}
\noindent In the centralized regime ($\mathbf{P_2}$), \eqref{eq:cons1} are the transmission AC power flow constraints from ~\eqref{eq_1} to \eqref{eq_7} with infeasibility sources, and $\Omega$ is the positive sequence component.
\eqref{eq:cons2} are the three-phase distribution power flow constraints (also in \eqref{eq_1} to \eqref{eq_7}) with $\Omega \in \{a,b,c\}$.
The infeasible sources are added to all or a subset of T\&D buses except the slack bus to provide the necessary mismatch in the AC network constraints. Depending on the choice of infeasibility source type (see Table \ref{tab:1}), the contribution of these sources to the network KCL equations in ~\eqref{eq_1} to \eqref{eq_7} can be linear (in case of currents and impedance) or nonlinear (in case of power).
% \vspace{-0.5em}
The constraint set highlighted by \eqref{eq:cons3} and \eqref{eq:cons4} are the physical and stability bounds on node voltages and branch flow in the T\&D sub-circuits. 
The coupling constraints in \eqref{eq:cons5} (derived from \eqref{C_eq_1} and \eqref{couple_eq_2}) connect the transmission and distribution system at the POI.
The objective minimizes the norm of the infeasibility terms $\mathcal{I}^T$ and $\mathcal{I}^D$.
The choice of norm (see $p$ in \eqref{eq:objective}) results in different behavior by each infeasibility source to satisfy AC network constraints. While the norm-1 minimization results in sparse non-zero infeasibility terms, norm-2 spreads the non-zero values across the majority of infeasibility sources \cite{boyd2004convex}.
\begin{subequations} \label{eq:centralized}
\begin{alignat}{3}
    & \mathbf{P}_{2}: \min_{X,\mathcal{I}}  \sum_{s \in S^T} \lVert(\mathcal{I}^{T}_{s})\rVert_{p}^{p} +  \sum_{s \in S^D} \sum_{\Omega \in {a,b,c}} \lVert(\mathcal{I}^{D}_{s, \Omega})\rVert_{p}^{p}\label{eq:objective} \\
    &\textrm{\text{subject to:}} \nonumber \\
%\end{alignat}
%\begin{alignat}{3}
    &\mathcal{F}^{T}_{s_t}(X_{s_t}^{T}) - \mathcal{I}^{T}_{s_t}  = 0 \quad \forall s_t \in S^T \label{eq:cons1} \\
    &\mathcal{F}^{D}_{s_d, \Omega}(X_{s_d,\Omega}^{D}) - \mathcal{I}^{D}_{s_d, \Omega}  = 0 \Hquad \forall \Omega \in \{a,b,c\} \Hquad \forall s_d \in S^D \label{eq:cons2} \\
    &\mathcal{G}^{{T}}_{s_t}(X^{T}_{s_t}) \leq 0 \quad \forall  s_t \in S^T \label{eq:cons3} \\
    &\mathcal{G}^{D}_{{s_d}, \Omega}(X^{D}_{s_d,\Omega}) \leq 0 \quad \forall \Omega \in \{a,b,c\} \Hquad \forall s_d \in S^D \label{eq:cons4} \\
    &\mathcal{C}_{k} (X^{{T}},X^{{D}}) = 0 \quad \forall k \in K \label{eq:cons5}
\end{alignat}
\end{subequations}
We use the perturbed primal-dual interior point method with Newton's steps to solve the centralized problem \cite{wachter2006implementation}.
If we minimize the norm-1 objective, we use the epigraph reformulation form \cite{boyd2004convex} to result in a differentiable formulation directly incorporable in nonlinear programming solvers.
Note that if the combined T\&D power flow problem is feasible, the optimization-based approach should converge to an equivalent power flow solution with all zero infeasibility terms, independent of the choice of the norm. The choice of infeasibility source type results in the following infeasibility current contribution:
\vspace{-0.5em}
\begin{equation}
\mathcal{I} = \begin{cases} 
I^{R,\inf}\!+\!jI^{I,\inf} & \text{if } \mathcal{I} = I^{\inf} \\
(P^{\inf}\!-\!jQ^{\inf})/(V^{R}\!-\!jV^{I}) & \text{if } \mathcal{I} = S^{\inf} \\
(G^{\inf}\!+\!jB^{\inf})(V^{R}\!+\!jV^{I}) & \text{if } \mathcal{I} = Z^{\inf}
\end{cases}
\end{equation}
\vspace{-0.2em}
\subsection{Distributed Approach}
\begin{figure}
    \centering
\includegraphics[width=\columnwidth]
{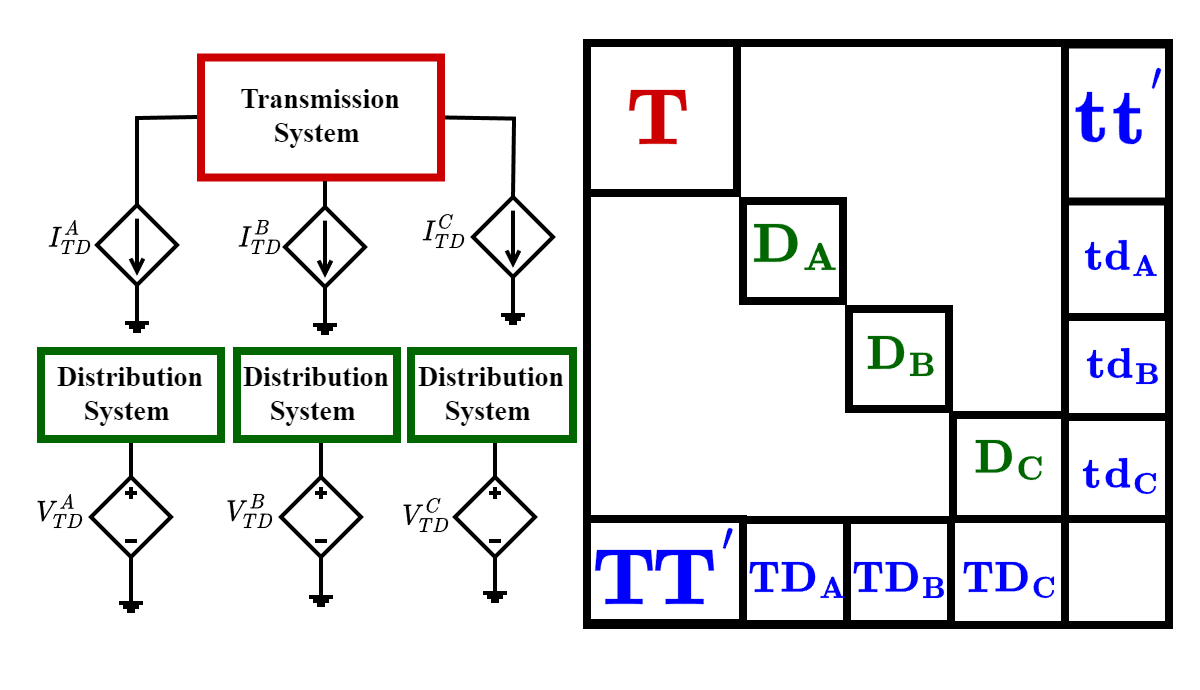}
     \caption{Combined T\&D in BBD structure. Here, A, B, and C represent three distinct distribution networks.}
    \label{fig:6}
\end{figure}

\vspace{-0.2em}
\noindent The centralized problem in \eqref{eq:objective}-\eqref{eq:cons5} may not scale indefinitely due to a single machine's limited compute capacity and memory buffer. 
Moreover, privacy limitations also prevent using a centralized approach because separate T\&D entities are hesitant to share full internal network data with each other. 
Given natural weak-coupling between T\&D networks, a distributed optimization approach is well-suited. 
In power system studies, numerous problems are addressed using distributed optimization; see \cite{molzahn2017survey}, \cite{dall2013distributed}.  
However, most of this research does not directly apply to \iv{}-based combined T\&D infeasibility analysis. Alternate direction method of multiplier (ADMM) and variants that behave closer to first-order methods are common choices for distributed computations in loosely coupled problems, as noted in \cite{annergren2014distributed}. 
These methods benefit from simplicity in local computations and convergence guarantees in convex setups, but they are limited in performance and convergence robustness in non-convex setups \cite{khoshfetrat2015distributed}. 
Furthermore, in ADMM-like methods that use first-order updates for all duals, iterations may increase significantly when accuracy demands are high, resulting in approximate solutions that may negatively affect the algorithm's convergence. We observe this behavior in the Section \ref{Experiments}. We implement a distributed primal dual interior point method to overcome these challenges to solve the combined T\&D problem\cite{khoshfetrat2015distributed}.
It differs from ADMM as we use the dual function's second-order information with Newton's step to update the majority of dual variables within the boundary of each subproblem.
We apply a Gauss-like step only for a small subset of dual variables, specifically those that couple the different subproblems (corresponding to the dual variables of coupling constraint in \eqref{eq:cons5}).
To partition the network, we apply domain-based decomposition, which partitions the original problem into multiple sub-problems $s \in S$ by branch tearing technique across the T\&D interface.
For efficient performance of this decomposition, we state the Remark 1.
The result is that the underlying solution matrix of perturbed primal-dual KKT conditions for the decomposed problem has a bordered block diagonal (BBD) form as shown in Fig. (\ref{fig:6}) \cite{pandey2024combined}.

\noindent \textbf{Remark 1.} \textit{The BBD structure depicted in Fig. (\ref{fig:6}) satisfies a key requirement: the dimension of the external variables ($X^{ext}_s$) is much smaller than the dimension of the internal variables ($X^{int}_{s}$).}

\noindent In the decomposed regime, the node tearing results in two sets of variables for each sub-problem: internal $(int)$ and external $(ext)$. 
Internal variables represent relationships between states of each T or D subnetwork, internally. 
External states appear in the coupling constraints and capture the relationship between two sub-problems.
Visually, in Fig. (\ref{fig:6}), the diagonal terms ($\mathbf{T}$, $\mathbf{D_{A}}$, $\mathbf{D_{B}}$, $\mathbf{D_{C}}$) are the function of internal primal and dual variables for each sub-problem. 
The off-diagonal entries ($tt', td_{A}, td_{B}, td_{C}$) are the function of external primal and dual variables with the entries at the bottom of the BBD structure ($TT', TD_{A}, TD_{B}, TD_{c}$) map constraints of the coupling circuit. 
The advantage of representing the problem in BBD form lies in its ability to leverage parallel computation by solving each diagonal block independently and using a sequential step to satisfy the relationship in the off-diagonal terms. The decomposed problem is given in $\mathbf{P_3}$ \eqref{eq:kkt_deco} with the external variables capturing the relationships of off-diagonal entries.
In the decomposed perturbed KKT constraints, equations \eqref{eq:sub1} and \eqref{eq:sub2} represent the primal and dual problems, respectively, for all subsystems. Equation \eqref{eq:sub3} denotes the perturbed complementary slackness condition, where $\mu_{s}$ is the vector of Lagrangian multiplier for the inequality constraints in sub-system $s$, and $\epsilon$ represents an epsilon small tolerance value. 
The equations \eqref{eq:sub4}-\eqref{eq:sub5} correspond to the dual and primal feasibility, respectively. We solve the decomposed PDIP problem with the Gauss-Jacobi-Newton algorithm.
\begin{subequations} \label{eq:kkt_deco}
\begin{align}
& \mathbf{P}_{3}: \text{Decomposed KKT conditions} \quad \forall s \in \{S^T, S^D\} \notag \\
&\nabla_{\lambda,s}\mathcal{L} = \mathcal{F}_{s}(X^{int}_{s}, X^{ext}_{k,s}) - \mathcal{I}_s^{int} = 0 \label{eq:sub1} \\
&\nabla_{x,s}\mathcal{L} = \nabla_{x,s}(||\mathcal{I}_s||_{p}^{p}) - {{\nabla_{x,s}} \mathcal{G}_{s}}^T \mu_{s}^{int} + \notag \\
&{{\nabla_{x,s}}(\mathcal{F}_{s}(X_s^{int}, X_{k,s}^{ext}}, \lambda_{k,s}^{ext}) - \mathcal{I}_s^{int})^T \lambda_s^{int}  = 0 \label{eq:sub2} \\
&-\mu_{s}^{int} \mathcal{G}_{s}(X^{int}) + \epsilon = 0\label{eq:sub3} \\
& \mu_{s}^{int} \geq 0 \label{eq:sub4} \\
& \mathcal{G}_{s}(X^{int}) \leq 0  \label{eq:sub5} 
\end{align}
\end{subequations}
We describe it in Algorithm \ref{Algm_1}. 
Within each epoch, independent sub-problems are optimized by solving their perturbed KKT conditions in parallel with Newton's method. The independent optimizations are terminated upon convergence or after a predetermined number of iterations. The external variables in each sub-problem $s$ are assumed fixed for the Newton problem. This step is equivalent to solving the block diagonal part of perturbed KKT conditions for each sub-problem $s$ with Newton's method while keeping the off-diagonal external terms fixed. 
We update the primal external variables ($X^{ext}_{k,s}$) for both transmission and distribution (T\&D) networks via coupling port relationships in \eqref{C_eq_1}-\eqref{couple_eq_2} with a Gauss-step.
\noindent To update the external dual variables at the coupling ports, we analyze the KKT conditions of $\mathbf{P_2}$. After some \href{https://github.com/HamZaAlipak/Combined-T_D_test_cases}{algebraic manipulation} of its KKT conditions, we derive the relationship between the dual variables of the T\&D coupling buses in \eqref{dual_expr}. We use this relationship to apply the Gauss step for dual variables.
In \eqref{dual_expr}, $\lambda^{D}_{k,\Omega}$ and $\lambda^{T}_k$ represent the dual variables for the distribution and transmission coupling buses, respectively. 
The algorithm is terminated when the exchange of information between two subsequent epochs is less than the defined tolerance $||X^{ext}_{k, n} - X^{ext}_{k, n-1}|| \leq \epsilon$. With this approach, entities only need to communicate minimal boundary primal and dual states between each other, which preserves privacy. 
The convergence of GJN depends on the conditions necessary for both Newton and Gauss steps. We discuss the general properties in Appendix A.
%\vspace{-1.1em}
\begin{equation}
    \begin{bmatrix}\label{dual_expr}
        \lambda^{R,D}_{k,a} \\
        \lambda^{I,D}_{k,a} \\
        \lambda^{R,D}_{k,b} \\
        \lambda^{I,D}_{k,b} \\
        \lambda^{R,D}_{k,c} \\
        \lambda^{I,D}_{k,c}
    \end{bmatrix} = 
    \cfrac{1}{\kappa}
        \begin{bmatrix}
             1 & 0 \\
             0 & 1 \\
             \alpha^{2} & 0 \\
             0 & \alpha^{2} \\
             \alpha & 0 \\
             0 & \alpha 
         \end{bmatrix}
    \begin{bmatrix}
        \lambda^{R,T}_{k} \\
        \lambda^{I,T}_{k}
    \end{bmatrix}
\end{equation}
\vspace{-1.0em}
\begin{algorithm}[htbp]
\caption{{\bf Distributed primal dual\\ 
interior point (D-PDIP)} \label{Algm_1}}
    \SetKwInOut{Input}{Input}
    \SetKwInOut{Output}{Output}
    \Input{T\&D network models, coupling nodes, infeasibility type and locations}
    \Output{T\&D grid infeasible locations}
    \textbf{Read T\&D input files}\\
    \textbf{Set} epoch $n=0$\\
    \textbf{Initialize} $X_s^{int}, \lambda_s^{int} \quad \forall s \in \{S^T, S^D\}$
    \\
     \For{$n=1$ \textbf{to} $N$}{
        \For{each sub-system $s$ in parallel}{
        \If{n=1}
        {
        \textbf{Initialize} coupling port variables $X_{k,s}^{ext}$, $\lambda_{k,s}^{ext}\Hquad \forall s \in \{S^T, S^D\}$ 
        }
        \textbf{Solve} the KKT conditions in \eqref{eq:sub1}-\eqref{eq:sub3} to convergence or predetermined steps with PDIP\\
        \textbf{Store} the $X^{int}_s, \lambda^{int}_s$ of coupling nodes for Gauss-step and others for warm-starting the next epoch
        }
    \textbf {Calculate/Update} $X^{ext}_{k,s}$ and $\lambda^{ext}_{k,s}   \forall s \in \{S^T, S^D\}$ using \eqref{C_eq_1},\eqref{couple_eq_2} and \eqref{dual_expr}\\
    \textbf{Check} convergence and break if tolerance met, $||y_{n} - y_{n-1}|| \leq \epsilon, y \in \{X^{ext}_{k}, \lambda^{ext}_{k}\}$
    }
\textbf{Report} results to grid planners
\end{algorithm}
\section{Experiments}\label{Experiments}
\vspace{-0.5em}
\noindent To test the performance of the proposed approach, we run multiple experiments with different choices of infeasibility sources (see Table \ref{tab:1}) and varying network sizes.
The ability to choose from different infeasibility source types allows the user to explore different remedial actions. 
The choice of current $I^{\inf}$ as infeasibility type can resolve mismatch in the combined T\&D network that lacks a power flow solution, as it identifies discrepancies within the power flow equations. 
However, this information may not suffice for system planners, as actionable preventive measures may be necessary to counteract the infeasibility in the combined T\&D networks. 
Hence, we should identify and localize with \textit{physical} sources of mismatch, such as by using $P^{\inf}$ or $Q^{\inf}$ sources to provide practical guidance to grid planners. 
These parameters will translate to meaningful corrective actions, such as determining the load curtailment required to render the system feasible and identifying locations for installing the reactive power compensating devices.
\subsection{Case Description}
\vspace{-0.5em}
\noindent We study six combined T\&D networks in the experiments with the case summary described in Table \ref{tab:case_summary}. The case data are available at \url{https://github.com/HamZaAlipak/Combined-T_D_test_cases.git}.
For cases 1 through 4, we use $Q^{\inf}$ infeasibility source type to identify locations within the combined T\&D network where system planners can strategically deploy reactive power-compensating devices to alleviate challenges caused by voltage violations. 
In Cases 5 and 6, we analyze a larger infeasible combined T\&D network. Specifically, Case 6 utilizes real distribution data and selects the $I^{\inf}$ infeasibility source type for analysis. All simulations were conducted on an AMD Ryzen 9 computer with 32-GB RAM. The Ipopt solver (v13.14.10) was used for inner optimization tasks, while Python scripts handled Gauss and other calculations.
\vspace{-2pt}
\begin{table}[htpb]
    \centering
    \caption{Summary of the test cases}
    \label{tab:case_summary}
    \begin{threeparttable}
    \begin{tabularx}{\columnwidth}{|c|>{\centering\arraybackslash}X|}
    \hline
    \textbf{Name}  &  \textbf{T\&D test case} \\
    \hline
    Case 1 & 14-bus (T) + 4-bus (D) \\
    \hline
    Case 2 & 118-bus (T) + GC-12.47.1 (D) \\
    \hline
    Case 3 & PEGAS2869 (T) + GC-12.47.1 (D)\tnote{1} \\
    \hline
    Case 4 & ACTIVsg25k (T) + D-net\tnote{2} \\
    \hline
    Case 5 &  ACTIVsg70k (T) + D-net\tnote{2} \\
    \hline
    Case 6 &  ACTIVsg25k (T) + VEC-net\tnote{3} \\
    \hline
    \end{tabularx}
    \begin{tablenotes}
      \small
      \item[1] Taxonomy feeder, GC 12.47.1 with 36 three-phase nodes (108 single-phase nodes).
      \item[2] Synthetic urban meshed network with 1420 three-phase nodes (4260 single-phase nodes).
      \item[3] Vermont distribution network (8.8k three phase nodes) 
    \end{tablenotes}
    \end{threeparttable}
\end{table}
\subsection{Reactive Power Compensation of Infeasible Combined T\&D Networks}
\vspace{-0.5em}
\noindent We study reactive power compensation in  \textbf{Cases 1 through 4}. To render the original combined T\&D networks infeasible, we increased the load magnitudes on the distribution networks by a factor of 1.5 and enforced voltage bounds at system nodes, including the POI. We run infeasibility analyses with $Q^{\inf}$ sources at a subset of defined network locations to identify optimal locations for reactive power compensation. From the analysis, we identify the \textit{need for additional} reactive power with the total amount of 0.218 (pu), 0.84 (pu), 1.485 (pu), and 0.46642 (pu) for all Cases 1 through 4, respectively. These results provide useful information to system planners for optimally deploying corrective devices. 
For instance, in \textbf{Case-4}, we identify that bus number 17293 needs additional reactive power to make the system feasible. 
We find that adding a reactive power compensating device, such as a synchronous var compensator, at the identified location can restore the combined T\&D network's feasibility. 
\begin{table}[htpb]
    \centering
    \caption{Comparison of Infeasible Combined T\&D networks with ADMM, D-PDIP, and C-PDIP}
    \label{tab:ADMMvsPDIP}
    \begin{tabularx}{\columnwidth}{|c|c|c|c|>{\centering\arraybackslash}X|c|}
    \hline
    \textbf{Algorithm} & \textbf{Network} & \textbf{OF (p.u)} & \textbf{Iter.} & \textbf{Time(s)} \\
    \hline
    ADMM & Case-1 & 0.023 & 38 & 0.47 \\
    \hline
    D-PDIP & Case-1 & 0.023 & 21 & 0.44 \\
    \hline
    C-PDIP & Case-1 & 0.023 & 4 & 0.13 \\
    \hline
    ADMM & Case-2 & 0.352 & 71 & 1.27 \\
    \hline
    D-PDIP & Case-2 & 0.352 & 28 & 0.98 \\
    \hline
    C-PDIP & Case-2 & 0.352 & 15 & 0.44 \\
    \hline
    ADMM & Case-3 & -- & -- & -- \\
    \hline
    D-PDIP & Case-3 & 1.102 & 32 & 8.63 \\
    \hline
    C-PDIP & Case-3 & 1.103 & 16 & 1.37\\ 
    \hline
    ADMM & Case-4 & -- & -- & -- \\
    \hline
    D-PDIP & Case-4 & 0.108 & 88 & 102.04\\
    \hline
    C-PDIP & Case-4 & 0.108 & 83 & 65.71 \\
    \hline
\end{tabularx}
\begin{flushleft}
\small{Note: The OF is calculated by $\sum_{s \in S^T} \frac{1}{2} \lVert(\mathcal{Q}{^{T,\inf}})\rVert_{2}^{2}$}\\
\small{Note: the number of iterations represents the number of times the algorithm performs the backslash operator}. 
\end{flushleft}
\end{table}

\noindent \textbf{Comparison}:
\noindent To evaluate the performance of the proposed D-PDIP algorithm, we compare its results for \textbf{Cases 1 through 4} against the Alternating Direction Method of Multipliers (ADMM) method and Centralized PDIP (C-PDIP) as discussed in section \ref{opti}.We highlight the results in Table \ref{tab:ADMMvsPDIP}. For smaller cases, all three methods converge to the same solution. 
D-PDIP outperforms ADMM in terms of the number of iterations due to its utilization of second-order information for duals. Furthermore, for the larger cases, the ADMM fails to converge in a timely manner ($\leq$ 1800 sec.), whereas D-PDIP demonstrates robust convergence. 
C-PDIP achieves faster convergence than D-PDIP across all scenarios because the number of sub-networks is not sufficiently large. 
Therefore, the computing power of a single machine is sufficient.
Nonetheless, the distributed approach will be necessary to enforce privacy needs and solve larger networks.

\subsection{Analyzing Large Infeasible Combined T\&D Networks}
\begin{figure}[!b]
    \centering
    \begin{subfigure}[t]{0.45\textwidth}
        \centering
\includegraphics[width=\linewidth,height=0.9\linewidth,keepaspectratio]{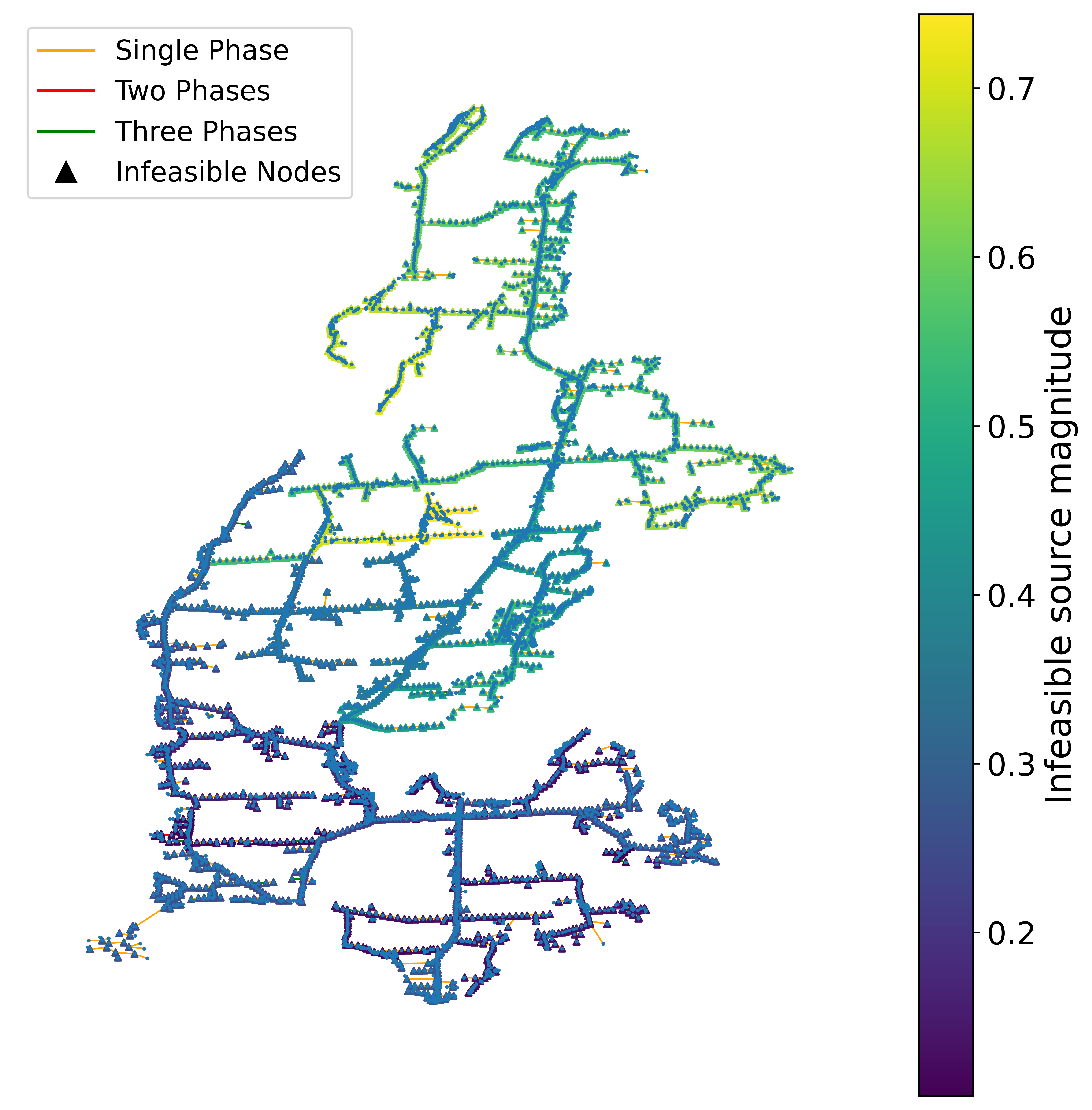}
        \caption{A}
        \label{fig:subfig1}
    \end{subfigure}
    \hfill
    \begin{subfigure}[t]{0.47\textwidth}
        \centering
\includegraphics[width=\linewidth,height=0.9\linewidth,keepaspectratio]{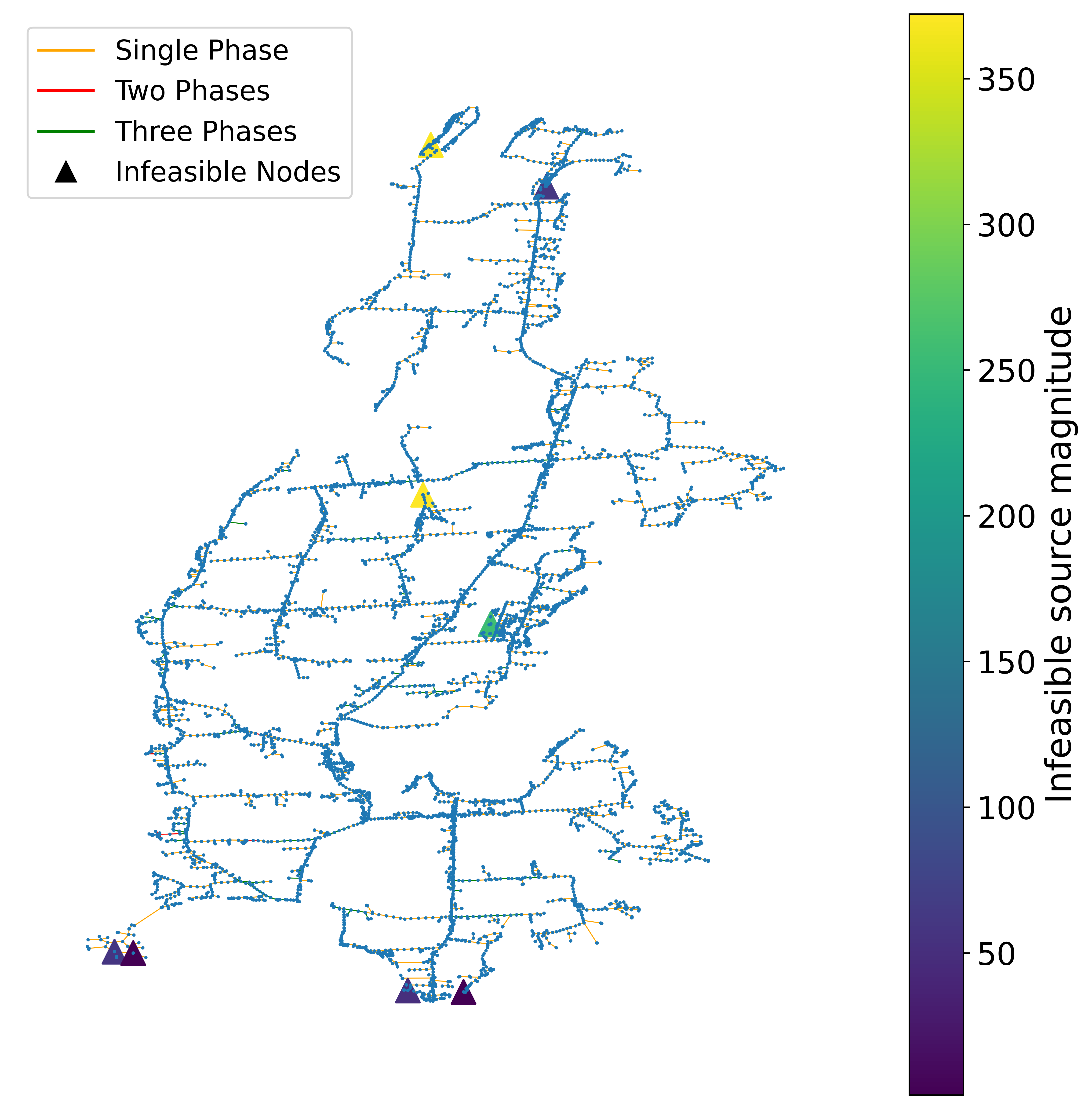}
        \caption{B}
        \label{fig:subfig2}
    \end{subfigure}
    \caption{A graph representation of the Vermont feeder with a heatmap showing the highest infeasibility current magnitude at each node, based on the L2 (A) and L1 (B) norm formulations. Note that with L2-norm, infeasibility currents are spread throughout the network. With L1 norm, non-zero infeasibility currents are only found in 8 locations.}
    \label{fig:L_L2}
\end{figure}

\vspace{-0.5em}
\noindent The \textbf{Case-6} network consists of a 25k node transmission network connected to a real distribution grid (substation in Vermont, which contains 8805 three phase nodes).
We render the original network infeasible by adding load to the system, expected due to electrification. 
We model the additional load with constant \textit{PQ} load model and we add it to a subset of triplex nodes in the system. 
Next, we select $I^{\inf}$ as an infeasibility source to solve this infeasible network with D-PDIP algorithm. 
The D-PDIP algorithm converges in three epochs, identifying various \textit{weak} locations in the transmission and distribution network. 
As illustrated in Figure \ref{fig:L_L2}, the infeasibility currents are distributed across almost 95\% of nodes.
This poses challenges for the system operator as no clear, actionable information exists. 
To mitigate this issue, we reformulate the optimization problem with \textit{L-1} norm, which can return a sparse vector of infeasibility currents. With infeasibilities in sparse locations, the system planner can apply the corrective actions to the few identified nodes to make the network feasible. 

\begin{figure}[htbp]
    \centering
\includegraphics[width=\columnwidth]{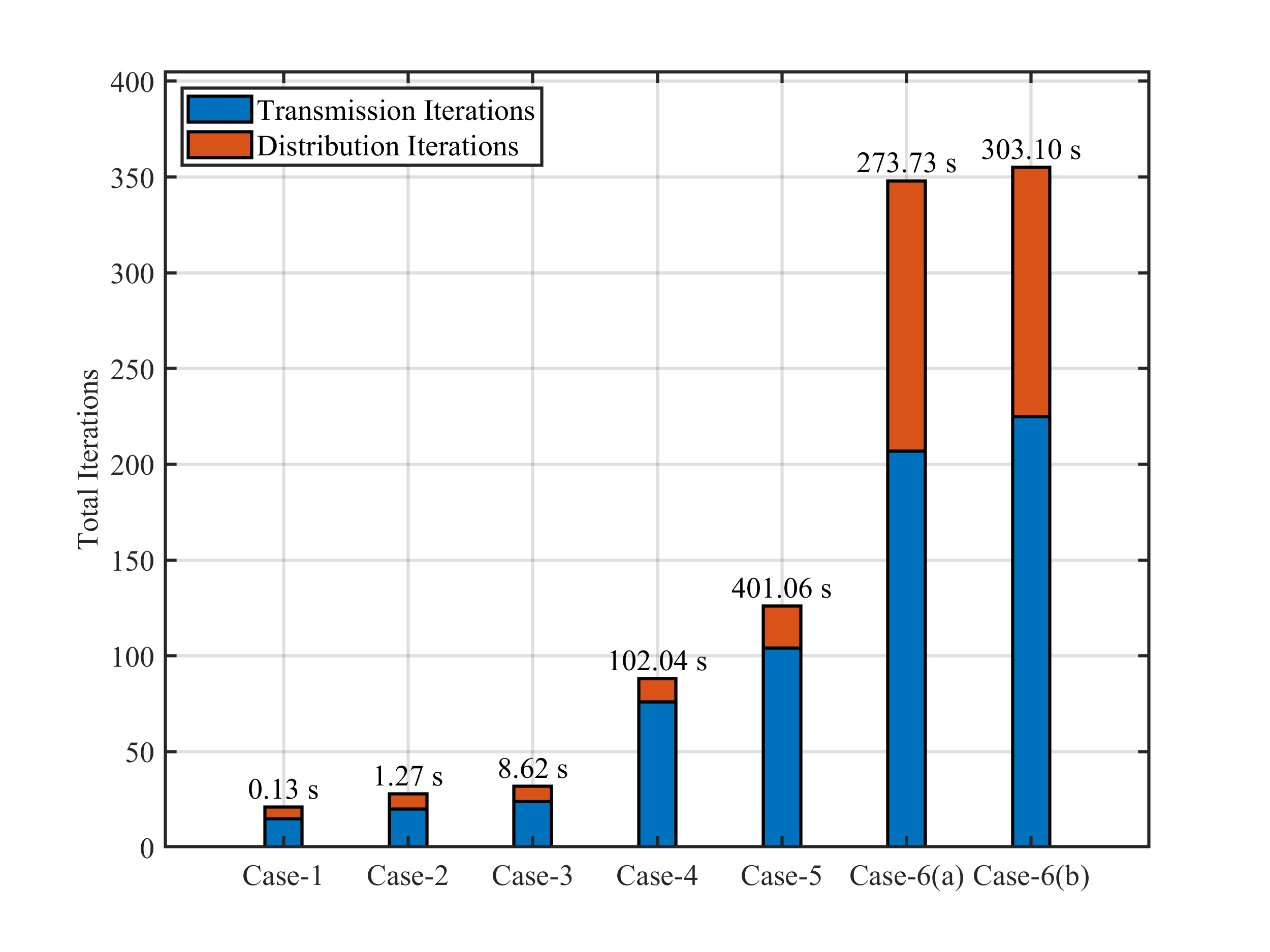}
     \caption{Combined T\&D D-PDIP algorithm scalability.}
    \label{fig:scal}
\end{figure}

\noindent \textbf{Scalability}:
Next, we demonstrate the scalability of the D-PDIP algorithm for \textbf{Cases 1 through 6} in Figure \ref{fig:scal}, the case-6 is divided into parts Case-6(a) and Case-6(b) representing the L-1 and L-2 formulation respectively. We present the total runtime and iterations required for the D-PDIP algorithm to converge, aggregating transmission and distribution iterations across epochs and subsystems. 
We show that the largest network with 70k+ transmission nodes and 1420 three-phase distribution nodes solves in less than 7 minutes.

\section{Conclusion}
\vspace{-0.5em}
\noindent In this paper, we present a novel distributed optimization framework to solve infeasible combined T\&D networks with actionable information. 
We model the joint T\&D networks in an equivalent circuit framework.
We use a circuit-based coupling port to couple the various T\&D sub-networks.
We apply node-tearning technique along the coupling ports to naturally decompose the problem.
We implement a distributed primal dual interior point (D-PDIP) algorithm to solve the decomposed infeasibility problem while i) enforcing privacy requirements and ii) scaling to combined networks greater than hundred thousand nodes.
Underneath, we solve the D-PDIP equations with the Gauss-Jacobi-Newton approach.
For solution robustness and fast convergence, we use second-order system information to solve the strongly coupled system states and a Gauss step to obtain consensus between weakly coupled primal and dual variables.
We show the approach's practical significance by identifying optimal reactive power compensation locations for large combined T\&D networks. Our approach achieves significantly better performance than the alternating direction method of multipliers (ADMM) approach.
We posit since this framework quantifies and localizes the deficiencies in combined T\&D networks, it can enable more effective planning and decision-making processes for grid operators and planners.
\section{Acknowledgement}
\vspace{-0.5em}
\noindent This research is funded by the National Science Foundation through contract ECCS: 2330195. 
%information of the coupling port within distributed optimization technique is utilized to expedite the convergence of the combined T\&D networks. We believe that our proposed framework holds significant practical applications in future electric grid planning with high penetration of DER. Since this framework quantifies and localizes the deficiencies associated with active and reactive power in a combined T\&D networks, which enables more effective planning and decision-making processes for grid operators and planners. 
%Future work will focus on considering the effect of inverter based resources (IBR) in the distribution side, and how it will effect the overall performance of proposed distributed optimization framework.  
%\bibliography
\printbibliography
% \bibliographystyle{ieeetr}
% \bibliography{sample}
\section*{Appendix-A}\label{Aappendix}
\vspace{-0.7em}
\noindent Consider a general optimization problem for subsystem $s$ in Algorithm \ref{Algm_1} with external variables fixed:
\begin{subequations}\label{appendix:eq1}
\begin{alignat}{3}
    &\min_{X} f (X_s) \\
    &\text{s.t.} \nonumber \\
    & H(X_s) = 0 \\
    & G(X_s) \leq 0 
\end{alignat}
\end{subequations}
The Lagrangian for the problem \eqref{appendix:eq1} is:
\begin{equation}
    \mathcal{L}(X_s,\lambda_s, \mu_s) = f(X_s) + \lambda_s^{T} H(X_s) + \mu_s^{T}G(X_s)
    % \mathcal{L}_s(X,\lambda, \mu) = f_s(X) + \lambda_s^{T} H_s(X) + \mu_s^{T}G_s(X)
\end{equation}
Deriving the perturbed KKT conditions and linearizing the system of equations to solve iteratively with Newton step, we obtain the following for $i^{\textrm{th}}$ iteration at $n^{\textrm{th}}$ epoch:
\begin{align}
\label{appendix:eq2}
    &\underbrace{\begin{bmatrix}
        W_{s,i} & A_{s,i}^{T} & 1 \\
        A_{s,i} & 0 & 0 \\
        \mu_{s,i} & 0 & X_{s,i} \\
    \end{bmatrix}}_{\textstyle Y(X_{s,i}, \lambda_{s,i}, \mu_{s,i})}
    \underbrace{\begin{bmatrix}
        X_{s,i+1} \\
        \lambda_{s,i+1} \\
        \mu_{s,i+1}
    \end{bmatrix}}_{\textstyle V_{s,i+1}} \notag \\
    & \quad =
    \underbrace{\begin{bmatrix}
        -\nabla_{x}\mathcal{L}_{s,i} + W_{s,i}X_{s,i} + A^{s,i}\lambda_{s,i} + \mu_{s,i} \\
        -H(X_{s,i}) + A_{s,i}X_{s,i} \\
        \mu_{s,i} \cdot X_{s,i} - \epsilon
    \end{bmatrix}}_{\textstyle J(X_{s, i},\lambda_{s, i}, \mu_{s, i})}
\end{align}
The term $W_{s,i} = \nabla_{XX}^{2}\mathcal{L}_{s,i}$ and $A = \nabla_{X} H_{s,i}$. 
The $X_s$ corresponds to the internal states of subsystem $s$.
%and external constant element at the off-diagonal of BBD form at $i^{th}$ iteration and coupling point $k$. 
Similarly, $\lambda_s$ are the internal duals of subsystem $s$. 
The equation (\ref{appendix:eq2}) corresponds to a single Newton-Raphson iteration of sub-problem $S$ toward solving the nonlinear perturbed KKT conditions. 

%At convergence, the following relationship holds:
%\begin{equation}
%    Y^{*}_{s,i} V^{*}_{s,i+1} = J^{*}_{s,i}
%\end{equation}
For a Newton-Raphson (NR) to converge for each sub-system $s$, the initial guess should be close to nonlinear problem root $V_s^*$, and all functions and their derivatives should have Lipschitz continuity in the close neighborhood of $V_s^{*}$ \cite{kelley1999iterative}. 

Similarly, for the Gauss step that updates the external terms between subsystems $s \in S$, the condition is on the overall system matrix $Y^*$ that concatenates, after each epoch, individual system matrices obtained from the NR solution of various sub-problems ($Y_s^{*}, \forall s \in S$) along with off-diagonal external terms.
%can be represented as:
%\begin{equation}
%    (\hat{Y}^{i+1})V^{n} = J^{n} 
%    \label{gauss-step}
%end{equation}
To stipulate the conditions for convergence of the Gauss-step, we split the matrix $Y^*$ into two components (${Y}^* = M - N$) \cite{minot2015fully}. 
The sufficient condition then for the Gauss-Jacobi method to converge is that the spectral radius $r$ of decomposed matrix $Y^*$ must be less than 1. 
\begin{equation}
    r = \rho(M^{-1}N) < 1
\end{equation}

\end{document}